\documentclass[aps,prl,reprint,superscriptaddress,nofootinbib,floatfix]{revtex4-2}

\usepackage[utf8]{inputenc}
\usepackage[T1]{fontenc}
\usepackage{amsmath,amssymb,bm,siunitx,graphicx}
\usepackage[
    colorlinks=true,
    citecolor=blue,
    linkcolor=blue,
    urlcolor=blue
]{hyperref}

\graphicspath{{figures/}}

\newcommand{\prlsection}[1]{%
    \par\addvspace{0.65\baselineskip}%
    \noindent\textit{#1.---}\hspace{0.35em}\ignorespaces
}

\newcommand{\ELIaff}{%
    ELI Beamlines Facility, Extreme Light Infrastructure ERIC,
    Za Radnic{\'i} 835, 252 41 Doln{\'i} B{\v{r}}e{\v{z}}any,
    Czech Republic
}

\newcommand{\CTUaff}{%
    Czech Technical University in Prague,
    Faculty of Nuclear Sciences and Physical Engineering,
    B{\v{r}}ehov{\'a} 7, 115 19 Prague 1, Czech Republic
}

\newcommand{\nnp}{n_{\mathrm{NP}}}

\begin{document}

\title{Laser-Driven Electron Stripping from Nanoparticles in Plasma}

\author{Alfred Mishi}
\email[Corresponding author: ]{alfred.mishi@eli-laser.eu}
\affiliation{\ELIaff}
\affiliation{\CTUaff}

\author{Marcel Lama{\v{c}}}
\affiliation{\ELIaff}

\author{Dominik {\v{C}}{\'a}p}
\affiliation{\ELIaff}
\affiliation{\CTUaff}

\author{Orsolya Morvai}
\affiliation{\ELIaff}
\affiliation{\CTUaff}

\author{{Vojt{\v{e}}ch Janota}}
\affiliation{\ELIaff}
\affiliation{\CTUaff}

\author{Romain Caye}
\affiliation{\ELIaff}

\author{Antonia Morabito}
\affiliation{\ELIaff}

\author{Jaroslav Nejdl}
\affiliation{\ELIaff}
\affiliation{\CTUaff}

\author{Jan Psikal}
\affiliation{\CTUaff}

\author{Uddhab Chaulagain}
\affiliation{\ELIaff}

\begin{abstract}
Laser wakefield accelerators use plasma wake waves to produce high-energy electron beams over short distances. Nanoparticles can trigger localized electron injection, with their size and density influencing the accelerated beam charge. However, the electron charge of a nanoparticle does not determine how much charge is stripped by the laser. Here we show that a quasistatic electrostatic model predicts laser-driven electron stripping from nanoparticles, in agreement with particle-in-cell simulations in vacuum and helium plasma. The model identifies a saturation density that increases with laser amplitude and decreases with nanoparticle radius, separating complete from partial stripping. Above this density, the stripped electron fraction decreases and the stripped charge approaches a finite limit. Multicycle pulses produce more stripped charge than the model predicts due to time-dependent electron dynamics. In helium plasma, plasma wake waves induce additional electron stripping. These results give an understanding of the scaling of laser-driven electron stripping from nanoparticles and could inform nanoparticle parameter selection for future nanoparticle-assisted laser wakefield accelerators.

\end{abstract}

\maketitle

\clubpenalty=10000
\widowpenalty=10000

\prlsection{Introduction}
Laser wakefield acceleration (LWFA) uses laser-driven plasma waves to accelerate electrons~\cite{Tajima1979,Lu2006,esarey2009,Mangles2004,Leemans2006}. Controlling electron injection is essential for producing reproducible electron beams, as it determines which electrons enter the accelerating region of the wake~\cite{Faure2006,Geddes2008,Gonsalves2011,Buck2013}. Nanoparticle-assisted injection offers a means of controlling this process through the localized charge-separation field of a laser-irradiated nanoparticle, which can perturb nearby plasma electrons and enable their trapping. Cho \textit{et al.}~\cite{Cho2018} proposed this injection scheme and developed an analytical description of how the nanoparticle potential influences single-particle motion and subsequent injection into the wakefield. Their PIC simulations demonstrated electron injection under conditions where the plasma wake alone did not produce self-injection. They also found that increasing the nanoparticle electron density initially increases the injected beam charge before reaching saturation, which they attributed to the balance between the laser ponderomotive potential and the nanoparticle potential.

Nanoparticle-assisted injection was later demonstrated experimentally by Aniculaesei \textit{et al.}~\cite{Aniculaesei2019} using laser-ablated aluminium nanoparticles in helium. Compared with self-injection, they obtained higher electron energies and reduced energy spread and divergence. Xu \textit{et al.}~\cite{xu2022} localized nanoparticle insertion to a region of about $100\,\mu\mathrm{m}$, separating control of injection from laser evolution. Their optimized conditions produced bunch charges above 300~pC with an injection probability above 90\%. Nanoparticle-assisted acceleration was eventually demonstrated at approximately 10~GeV over 10~cm by Aniculaesei \textit{et al.}~\cite{aniculaesei2024}. More recently, \v{S}p\'{a}dov\'{a} \textit{et al.}~\cite{Spadova2025} investigated how nanoparticle material and size influence the accelerated beam charge. Their simulations identified a regime beyond saturation of the nanoparticle field in which the beam charge depends on the number of atoms in the nanoparticle and not its electron density after ionization. They also found that injection into several plasma-wave periods increases the total beam charge but increases energy-spread.

These studies demonstrate the importance of nanoparticle properties in controlling electron injection and acceleration. However, the electron charge of a nanoparticle does not necessarily determine how much charge is stripped by the laser. Also, electrons stripped from the nanoparticle must be trapped in the accelerating phase of the plasma wake to contribute to the accelerated beam. Understanding the physics and dynamics of nanoparticle charge stripping by the laser in plasma is therefore an important step in determining the electrons available for possible electron injection and acceleration in the plasma wake.

The theoretical description of electron stripping from charged clusters provides a basis for studying this process in plasmas. Breizman and Arefiev~\cite{BreizmanArefiev2003} derived the equilibrium of a cold electron sphere displaced inside a uniform ion sphere. For electrons that escape without returning, they expressed the smallest unstripped electron volume to the maximum applied field. They also studied collective electron motion and heating associated with reflection at the cluster boundary. Breizman \textit{et al.}~\cite{Breizman2005} attributed the number of stripped electrons to the difference between the ion-sphere and unstripped-electron volumes. Their nonlinear theory distinguished a cold electron core from a stochastically heated electron population and studied how these electron populations affect ion expansion. Arndt~\cite{Arndt2006} expressed the outer-ionization fraction, defined as the fraction of electrons stripped from the cluster, in terms of the volume occupied by the remaining electrons. For a uniform spherical cluster of radius $R$ and a retained electron sphere of radius $r_u$, this gives $S=1-(r_u/R)^3$. We apply the same geometrical understanding to laser-driven electron stripping from nanoparticles.

The quasistatic approximation assumes that the retained electrons remain cold and follow the instantaneous electrostatic equilibrium within a fixed ion sphere. Our model adopts these assumptions and neglects electron inertia, heating, collective oscillations, electron return, and the evolving field of stripped electrons. These effects can influence the electrons remaining inside the nanoparticle and the stripped charge over successive optical cycles of the laser pulse~\cite{Fennel2010,Kull2018}. The model also does not account for additional electron stripping induced by the plasma wake wave after the laser-driven electron stripping. Studying these limitations is important for understanding when the electrostatic model can reliably predict the charge stripped from nanoparticles.

In this paper, we study an electrostatic model for laser-driven electron stripping from nanoparticles in vacuum and plasma. Building on the quasistatic theory of the retained electron volume, we derive a complete-stripping saturation density and a finite high-density limit for the stripped charge. We compare the predicted stripped electron fraction, charge, and laser amplitude and nanoparticle radius scaling with quasi-three-dimensional (Quasi-3D) PIC simulations. We then study the evolution of plasma wake wave and the timing of electron departures from the nanoparticle surface to identify additional stripping associated with the plasma wake wave. Finally, we investigate the dependence of electron stripping on laser pulse duration in vacuum to assess the influence of time-dependent electron dynamics and the limitations of the electrostatic model.

\prlsection{Model}
Consider an initially neutral nanoparticle of radius $R$, with uniform electron density $\nnp$ and fixed positive charge density $e\nnp$, where $e>0$ is the elementary charge. The total initial electron-charge magnitude is $Q_0=4\pi e\nnp R^3/3$. Here $\nnp$ is the initial nanoparticle electron number density; for an ion charge state $Z$, the corresponding ion density is $\nnp/Z$. In the following, we use the quasistatic theory of the unstripped electron volume developed in cluster theory~\cite{BreizmanArefiev2003,Breizman2005,Arndt2006}.

Suppose that the electrons remaining inside the ion sphere are cold and in electrostatic equilibrium under a spatially uniform applied field $\bm E$, with magnitude $E=|\bm E|$. Force balance requires $\bm E+\bm E_i+\bm E_{\mathrm{u}}=0$ throughout the region they occupy, where $\bm E_i$ and $\bm E_{\mathrm{u}}$ are the fields of the ions and unstripped electrons. Taking the divergence and applying Gauss's law gives $e(\nnp-n_{\mathrm{u}})/\epsilon_0=0$, because $\nabla\cdot\bm E=0$. Hence, the density $n_{\mathrm{u}}$ of the unstripped electrons equals $\nnp$. Electron stripping reduces the unstripped electron volume while this equilibrium density remains unchanged.

The unstripped electrons can be represented by a sphere of radius $r_{\mathrm{u}}$, whose center is displaced from the ion center by $\bm{\delta}_r$. At a point $\bm r$ inside both spheres, the uniform-sphere fields are $\bm E_i=e\nnp\bm r/(3\epsilon_0)$ and $\bm E_{\mathrm{u}}=-e\nnp(\bm r-\bm{\delta}_r)/(3\epsilon_0)$~\cite{FitzpatrickSphere}. Their position-dependent parts cancel, leaving a uniform restoring field. Force balance gives
\begin{equation}
    \begin{aligned}
        \bm E_i+\bm E_{\mathrm{u}}&=\frac{e\nnp}{3\epsilon_0}\bm{\delta}_r,\\
        \bm{\delta}_r&=-\frac{3\epsilon_0}{e\nnp}\bm E,
    \end{aligned}
    \label{eq:displacement}
\end{equation}
The unstripped electron sphere is displaced opposite to the applied electric field, with $\delta_r=|\bm{\delta}_r|=3\epsilon_0 E/(e\nnp)$.

The balance condition requires that $r_{\mathrm{u}}+\delta_r\le R$. In the quasistatic stripping theory, an initially neutral sphere subjected to a rising field loses electrons until the unstripped electron sphere is internally tangent to the ion sphere~\cite{BreizmanArefiev2003,Breizman2005}. For partial stripping, the radius of the unstripped sphere becomes $r_{\mathrm{u}}=R-\delta_r$, with $\delta_r<R$. Complete stripping is reached when $\delta_r=R$ and $r_{\mathrm{u}}=0$. This assumes a spherical unstripped electron volume and negligible field from electrons already stripped. If those electrons do not return as the field falls, the largest field reached during the pulse determines the smallest unstripped electron volume.

The number of unstripped electrons is $N_{\mathrm{u}}=4\pi\nnp r_{\mathrm{u}}^3/3$, whereas initially $N_0=4\pi\nnp R^3/3$. Hence the stripped fraction is $S=1-N_{\mathrm{u}}/N_0=1-(r_{\mathrm{u}}/R)^3$. The complete-stripping condition $\delta_r=R$ at peak applied field $E_0$ gives
\begin{equation}
    \begin{aligned}
        n_B&=\frac{3\epsilon_0 E_0}{eR}
             =\frac{3a_0n_c}{k_0R},\\
        S&=
        \begin{cases}
            1, & \nnp\le n_B,\\[3pt]
            1-\left(1-\dfrac{n_B}{\nnp}\right)^3,
            & \nnp>n_B,
        \end{cases}
    \end{aligned}
    \label{eq:model}
\end{equation}
where $a_0=eE_0/(m_ec\omega_0)$, $n_c=\epsilon_0m_e\omega_0^2/e^2$, and $k_0=\omega_0/c$. Here $n_B$ is the largest electron density for which the field $E_0$ completely strips a nanoparticle of radius $R$ within the model. The cubic dependence follows from the unstripped electron volume~\cite{Arndt2006}. For $\nnp\le n_B$, the model gives $S=1$; for $\nnp>n_B$, a finite unstripped electron sphere remains.

The charge stripped by the laser is $Q_{ne}=Q_0S$. For $\nnp>n_B$, expanding Eq.~\eqref{eq:model} gives
\begin{equation}
    \begin{aligned}
        Q_{ne}&=Q_\infty\left[1-\frac{n_B}{\nnp}
                     +\frac{1}{3}\left(\frac{n_B}{\nnp}\right)^2\right],\\
        Q_\infty&=12\pi\epsilon_0 E_0R^2,
    \end{aligned}
    \label{eq:charge}
\end{equation}
Increasing the density above $n_B$ reduces the stripped fraction while increasing the stripped charge toward the finite limit $Q_\infty$. At $\nnp=n_B$, the charge is only $Q_\infty/3$.

\prlsection{Simulations and methods}
\label{sec:simulations_methods}
We performed two sets of particle-in-cell (PIC) simulations using Smilei~\cite{Smilei2018}. The first comprises 450 simulations used to benchmark the electrostatic model and investigate laser-driven and plasma wake wave electron stripping. The second comprises 14 vacuum simulations used to study the influence of laser pulse duration and the time-dependent electron dynamics not currently included in the model. Since the nanoparticle is spherical and placed on the laser propagation axis, we use cylindrical coordinates with azimuthal mode decomposition, using modes $m=0,1$~\cite{Lifschitz2009}. These modes represent the cylindrically symmetric field components and the linearly polarized laser while reducing the computational cost compared with full three-dimensional cartesian simulations.

\begin{figure*}[!t]
    \centering
    \includegraphics[width=\textwidth,height=0.70\textheight,keepaspectratio]{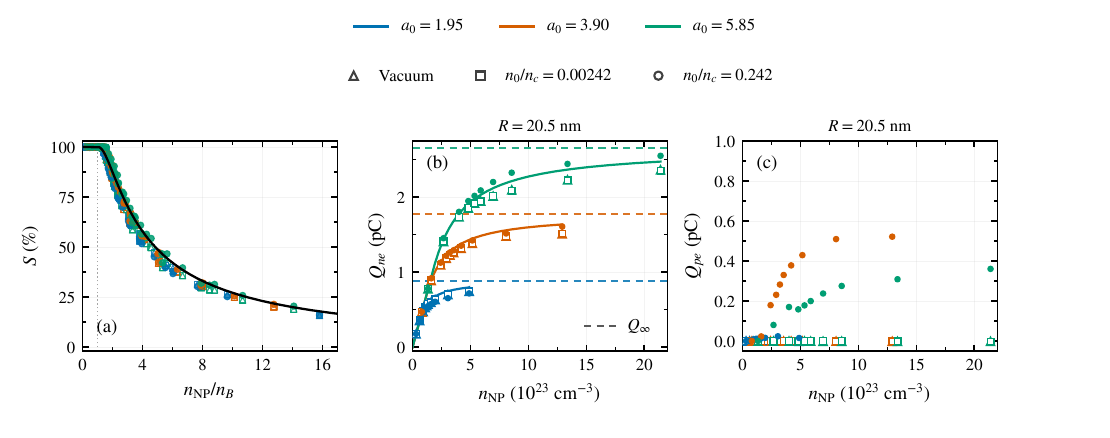}
    \caption{Nanoparticle electron stripping by the laser pulse and plasma wake wave. (a) Laser-driven stripped electron fraction; the black curve is Eq.~\eqref{eq:model}. (b) Laser-driven stripped charge $Q_{ne}$. (c) Plasma wake wave charge stripping $Q_{pe}$ at nanoparticle radius $R=20.5$~nm. Curves in (b) show the model prediction using the nanoparticle electron charge. Panel (c) contains PIC results only.}
    \label{fig:stripping}
\end{figure*}

For the first set of 450 benchmark simulations, we consider three laser amplitudes, $a_0=1.95$, $3.90$, and $5.85$, and five nanoparticle radii, $R=20.5$, $25.8$, $32.5$, $41.0$, and $51.7\,\mathrm{nm}$. For each amplitude--radius combination, ten nanoparticle electron densities are studied to cover complete and partial stripping, within a range from $4.81\times10^{21}$ to $2.14\times10^{24}\,\mathrm{cm}^{-3}$. The same 150 combinations are simulated in vacuum and in two helium plasmas with background electron plateau densities of $n_0=2.7\times10^{18}$ and $2.7\times10^{20}\,\mathrm{cm}^{-3}$. These correspond to $n_0/n_c=0.00242$ and $0.242$, respectively, where the critical density is $n_c=1.115\times10^{21}\,\mathrm{cm}^{-3}$ at $\lambda_0=1\,\mu\mathrm{m}$. Comparing the vacuum and helium plasma cases allows us to understand how the background plasma affects nanoparticle electron stripping. The laser propagates along $x$, is linearly polarized along $y$, and has a wavelength $\lambda_0=1\,\mu\mathrm{m}$ and Gaussian waist $w_0=\sqrt{3}\,\mu\mathrm{m}$. The waist is much larger than the nanoparticle, providing an approximately uniform transverse laser amplitude across it. The same Gaussian temporal envelope is used for all nanoparticle densities and laser amplitudes, with field- and intensity-envelope FWHM values of $\omega_0\tau_E=5.96$ and $\omega_0\tau_I=4.22$, respectively, where $\tau_I=\tau_E/\sqrt{2}$. The laser is focused at the end of a $3.33\,\mu\mathrm{m}$ sinusoidal-squared entrance ramp, which provides a smooth transition to the plasma plateau. The nanoparticle is centered on the propagation axis at $x=3.35\,\mu\mathrm{m}$, close to the laser focus. The nanoparticle is initialized as a charge-neutral sphere of cold electrons and fixed positive ions. Fixing the ions preserves the nanoparticle radius and ion distribution assumed by the model. The helium plasma is initially fully ionized and cold, with mobile ions. The smallest nanoparticle radius, $R=20.5\,\mathrm{nm}$, is used as the reference case in Fig.~\ref{fig:stripping}.

The second set (14 vacuum) PIC simulations consisted of seven matched configurations with intensity-FWHM durations of $\omega_0\tau_I=28.25$ and $65.93$ (at $\lambda_0=1\,\mu\mathrm{m}$). They were selected to vary nanoparticle electron density, radius, and laser amplitude while keeping other parameters constant. The selected densities are $n_{\mathrm{NP}}=(0.970,1.940,3.879)\times10^{23}\,\mathrm{cm}^{-3}$ at $R=32.54\,\mathrm{nm}$ and $a_0=2.3$; radii $R=20.5$, $32.54$, and $51.7\,\mathrm{nm}$ at $n_{\mathrm{NP}}=1.940\times10^{23}\,\mathrm{cm}^{-3}$ and $a_0=2.3$; and laser amplitudes $a_0=2.3$, $4.6$, and $6.9$ at $R=32.54\,\mathrm{nm}$ and $n_{\mathrm{NP}}=1.940\times10^{23}\,\mathrm{cm}^{-3}$.

The grid parameters for the two simulation sets are summarized in Table~\ref{tab:grid_parameters}. In the benchmark simulations, the grid spacing is adjusted according to nanoparticle radius. For the four smaller radii, the nanoparticle diameter have approximately 7 longitudinal and 8.4 transverse cell widths. The largest nanoparticle uses the same grid spacing as the $R=41.0\,\mathrm{nm}$ case, with 8.8 longitudinal cells across its diameter and 5.3 radial cells from the axis to its surface. The pulse-duration simulations use a stationary longitudinal--radial grid of $106.46\,\mu\mathrm{m}$ longitudinally and $54.98\,\mu\mathrm{m}$ radially. At $R=32.54\,\mathrm{nm}$, the nanoparticle is resolved with 11.1 longitudinal cells across its diameter and 6.7 radial cells from the axis to its surface. Each occupied nanoparticle cell contains 1000 macroparticles per species, and each helium plasma species has 8 macroparticles per cell.

\begin{table}[t]
\caption{Simulation grid parameters. Set I comprises the 450 benchmark simulations in vacuum and helium plasma; set II comprises the 14 pulse-duration simulations in vacuum. The grid dimensions $N_x\times N_r$ are the numbers of longitudinal and radial cells.}
\label{tab:grid_parameters}
\begin{ruledtabular}
\begin{tabular}{ccccc}
Sim. Set & $R$ (nm) & $N_x\times N_r$ & $\Delta x$ & $\Delta r$ \\
\hline
I & 20.5 & $2304\times960$ & $\lambda_0/171$ & $\lambda_0/205$ \\
I & 25.8 & $1920\times752$ & $\lambda_0/136$ & $\lambda_0/163$ \\
I & 32.5 & $1536\times608$ & $\lambda_0/108$ & $\lambda_0/129$ \\
I & 41.0 & $1152\times480$ & $\lambda_0/85$ & $\lambda_0/102$ \\
I & 51.7 & $1152\times480$ & $\lambda_0/85$ & $\lambda_0/102$ \\
II & 20.5 & $18176\times11264$ & $\lambda_0/171$ & $\lambda_0/205$ \\
II & 32.5 & $18176\times11264$ & $\lambda_0/171$ & $\lambda_0/205$ \\
II & 51.7 & $18176\times11264$ & $\lambda_0/171$ & $\lambda_0/205$ \\
\end{tabular}
\end{ruledtabular}
\end{table}

\begin{figure*}[!t]
    \centering
    \includegraphics[
        width=\textwidth,
        height=0.70\textheight,
        keepaspectratio
    ]{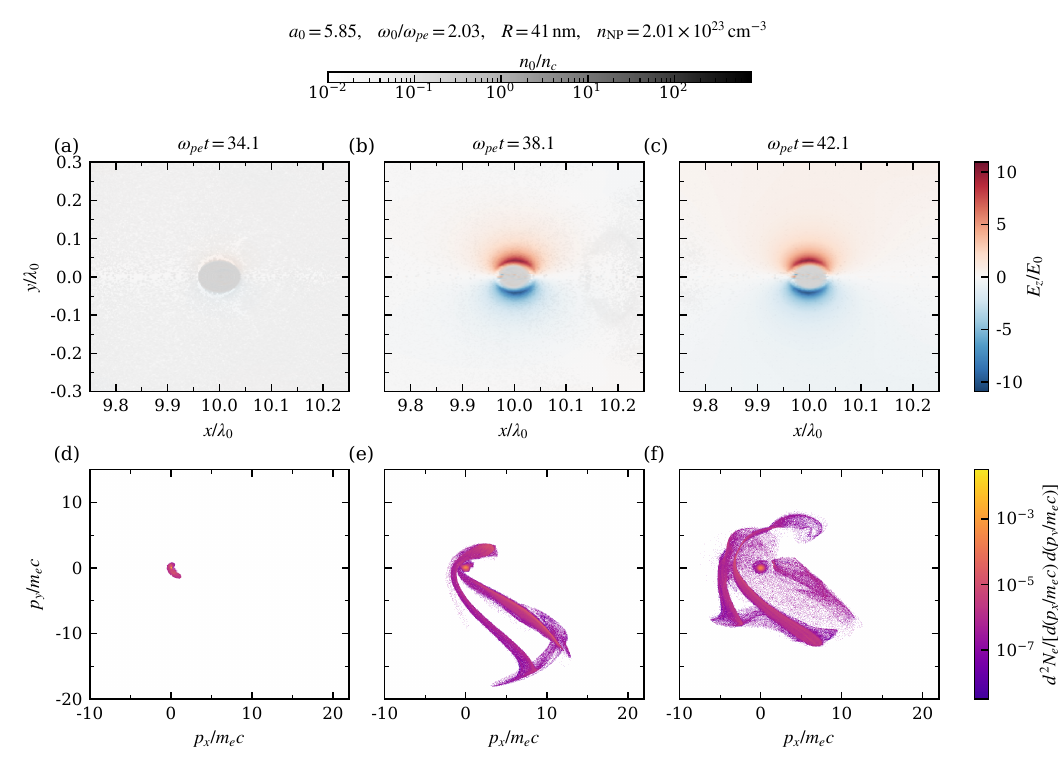}
    \caption{Nanoparticle charge separation, electron momentum, and field evolution in laser-nanoparticle interaction.
        (a)--(c) Growth of the nanoparticle charge-separation field due to laser-driven electron stripping relative to the fixed positive ions.
        (d)--(f) Electron momentum distributions of stripped electrons in the $(p_x,p_y)$ plane.}
    \label{fig:stripping_sequence}
\end{figure*}

\prlsection{Stripped electron fraction and charge}
To understand the physics underlying laser-driven electron stripping of nanoparticles in plasma, we first study the nanoparticle electron response to the applied laser field, as shown in Fig.~\ref{fig:stripping_sequence}. As the leading edge of the laser approaches and interacts with the nanoparticle, electrons are stripped relative to the fixed positive ion core, producing a growing charge-separation field that opposes their displacement. The initially weak field develops into a strong bipolar structure around the nanoparticle [Fig.~\ref{fig:stripping_sequence}(a)--(c)]. This evolution is accompanied by a broadening of the electron momentum distribution [Fig.~\ref{fig:stripping_sequence}(d)--(f)], with significant transverse momentum developing along the laser polarization direction.

We define the model electric field using the laser amplitude,
\begin{equation}
    a_0=\frac{eE_0}{m_ec\omega_0},\qquad
    E_0=a_0\frac{m_ec\omega_0}{e},
    \label{eq:laser_peak}
\end{equation}
where $m_e$ is the electron mass, $c$ is the speed of light in vacuum, and $\omega_0=2\pi c/\lambda_0$~\cite{Smilei2018}. At $\lambda_0=1\,\mu\mathrm{m}$, the three amplitudes give $E_0=6.26$, 12.52, and 18.78~TV/m, which are then used in Eqs.~\eqref{eq:model} and \eqref{eq:charge}. For the PIC comparison, let $Q_e(t)$ be the positive magnitude of the nanoparticle-electron charge enclosed within a fixed sphere of radius $R$. We define the enclosed charge as $Q_0=Q_e(t_0)$, with time, $t_0$ taken before the laser-nanoparticle interaction. The stripped electron charge follows as, $Q(t)=Q_0-Q_e(t)$. 
For comparisons between the electrostatic model and PIC simulations, we denote the laser-driven stripped charge as $Q_{ne}^{\mathrm{model}}=Q_0S$ and $Q_{ne}^{\mathrm{PIC}}$, respectively.

To quantify the plasma wake wave electron stripping observed in the $n_0/n_c=0.242$ helium plasma, we define
\begin{equation}
    Q_{pe}=\overline{Q}-Q_{ne},
    \label{eq:additional_charge}
\end{equation}
where $\overline{Q}$ is the time-averaged stripped charge over the interval $14\leq t-t_{\mathrm{ref}}\leq18$~fs, following the plasma wake wave interaction. Hence, $Q_{pe}$ includes the effect of electron return and measures the additional nanoparticle electron stripping associated with the plasma wake wave.

Figure~\ref{fig:stripping}(a) compares all 450 laser-driven stripped fractions with Eq.~\eqref{eq:model}. The normalization $\nnp/n_B$ uses the model prediction from the laser electric field amplitude. There are two regimes of nanoparticle electron stripping - complete and partial. The PIC simulations follow the transition from complete stripping regime to the partial stripping regime above the saturation density, $n_B$. Vacuum and lower-density helium plasma $n_{\mathrm{0}}/n_c=0.00242$ give nearly identical results. The denser-helium plasma $n_{\mathrm{0}}/n_c=0.242$ also follow the predicted saturation density and charge scaling.

We select a nanoparticle radius of $R=20.5\,\mathrm{nm}$ to show how the stripped charge varies with nanoparticle electron density for three laser amplitudes, $a_0=1.95$, $3.90$, and $5.85$, as shown in Fig.~\ref{fig:stripping}(b). Below the derived saturation density $n_B$, the stripped charge increases linearly with nanoparticle electron density. Above $n_B$, the increase becomes nonlinear and gradually approaches a finite limit. Increasing $\nnp$ increases the nanoparticle electron charge, but a larger fraction remains unstripped, so the increase in stripped charge per density increase becomes smaller. The colored curves show $Q_0S$ calculated using Eq.~\eqref{eq:model}. Figure~\ref{fig:stripping}(c) shows the additional charge stripped by the plasma wake wave, $Q_{pe}$, obtained from the PIC simulations. $Q_{pe}$ seems also to approach a finite limit/plateau charge, as observed in the laser-driven stripping

\begin{figure*}[!t]
    \centering
    \includegraphics[width=\textwidth]{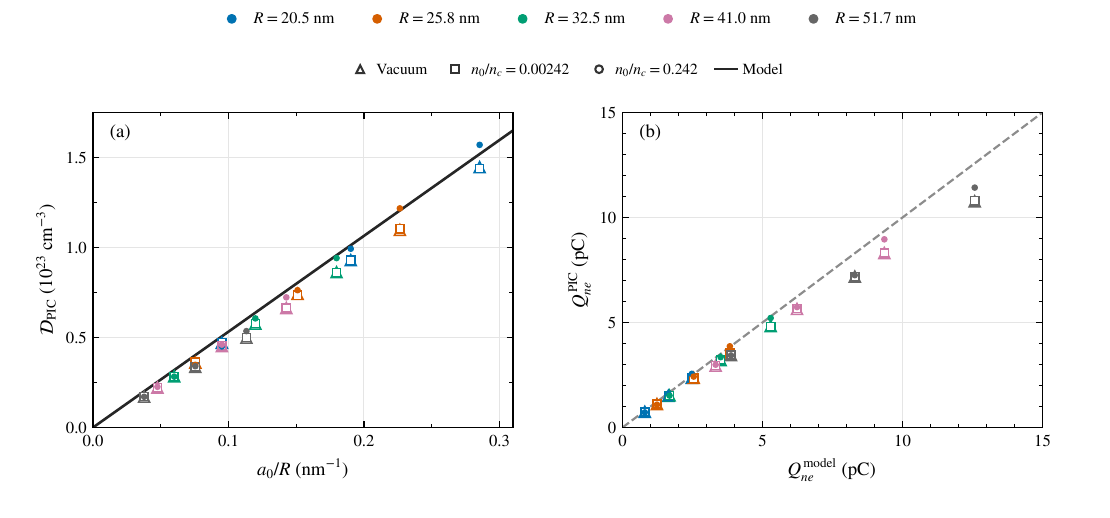}
    \caption{Laser amplitude and nanoparticle radius scaling of electron stripping.
    (a) Effective density $\mathcal{D}_{\mathrm{PIC}}$
    showing agreement with the model predicted linear dependence
    on $a_0/R$ (solid line).
    (b) Comparison of PIC measured and model predicted stripped charge.}
    \label{fig:scaling}
\end{figure*}

\prlsection{Laser amplitude and radius scaling}
The saturation density derived in Eq.~\eqref{eq:model} also describes how laser amplitude and nanoparticle radius influence electron stripping. To investigate this scaling correctly, we select 45 PIC simulations covering five nanoparticle radii and three laser amplitudes in vacuum and helium plasma. For each radius and laser amplitude, we select the highest nanoparticle electron density studied to focus on the partial-stripping regime, where the predicted saturation density can be directly tested against the PIC results. A stronger laser can completely strip electrons from a nanoparticle at higher electron densities, but a larger nanoparticle requires a stronger laser field to overcome the electrostatic force restoring electrons to the core. The model captures this scaling through $n_B\propto a_0/R$. To test this scaling in the PIC simulations, we define an effective density using the PIC stripped electron fraction $S_{\mathrm{PIC}}$, taken as $Q(t)/Q_0$ over the laser-driven stripping region. The effective density is then given by

\begin{equation}
\mathcal{D}_{\mathrm{PIC}}
=\nnp\left[1-\left(1-S_{\mathrm{PIC}}\right)^{1/3}\right].
\end{equation}

For partial stripping, Eq.~\eqref{eq:model} predicts

\begin{equation}
\mathcal{D}_{\mathrm{PIC}}=n_B
=\frac{3\epsilon_0m_ec\omega_0}{e^2}\frac{a_0}{R}.
\end{equation}

The results for the five nanoparticle radii and three laser amplitudes are in good agreement with the predicted linear dependence on $a_0/R$ in vacuum and helium plasmas [Fig.~\ref{fig:scaling}(a)]. This agreement shows that the model captures the combined scaling of electron stripping on laser amplitude and nanoparticle radius across the studied simulations. Figure~\ref{fig:scaling}(b), shows the predicted stripped charge of the model with the PIC results for the same 45 simulations. The predicted charges are in good agreement with the PIC results, although some deviations are observed at higher stripped charges. This comparison also highlights the distinction between the stripped electron fraction and the total stripped charge. Complete stripping does not necessarily mean that more charge is released. At the same electron density, a smaller nanoparticle is easier to completely strip but contains fewer electrons, while a larger nanoparticle can release more charge even when only part of its electrons are stripped.

\begin{figure*}[!t]
    \centering
    \includegraphics[
        width=\textwidth,
        height=0.70\textheight,
        keepaspectratio
    ]{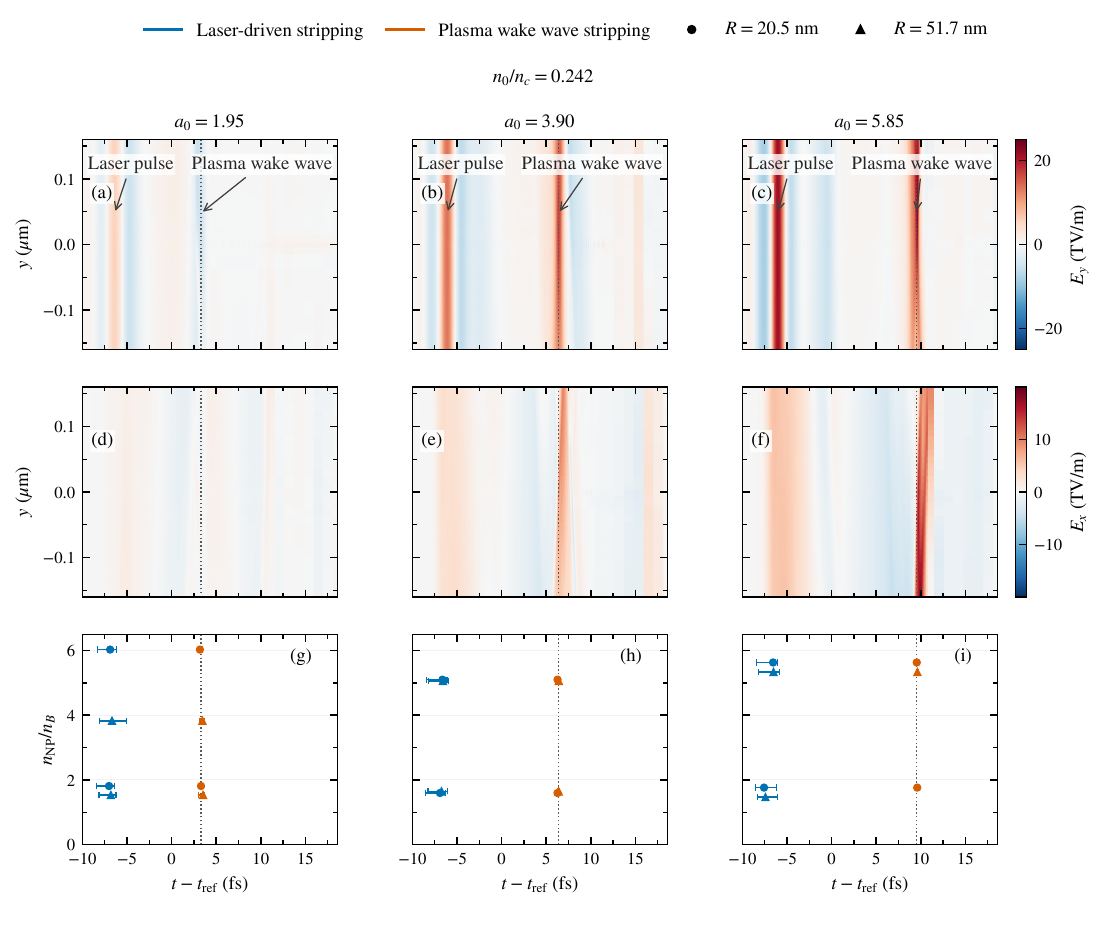}
    \caption{Laser-driven and plasma wake wave induced electron stripping from nanoparticles in helium plasma.
    (a)--(c) Evolution of the transverse electric field $E_y$ showing the laser pulse followed by the plasma wake wave.
    (d)--(f) Evolution of the longitudinal electric field $E_x$ associated with the plasma wake wave.
    (g)--(i) Timing of the first outward electron departures from the nanoparticle surface during the laser-driven and plasma wake wave interactions.}
    \label{fig:fields}
\end{figure*}

\prlsection{Helium plasma effects}
In the $n_0/n_c=0.242$ helium plasmas, the plasma wake wave induces additional electron stripping from the nanoparticles after the laser-driven stripping. To investigate this process, Fig.~\ref{fig:fields} compares the timing of stripped electron trajectories obtained from 12 selected cases with the electric fields calculated in separate simulations without nanoparticles. A common reference time, $t_{\mathrm{ref}}=21.1257\,\mathrm{fs}$, is chosen after the laser pulse and before the arrival of the plasma wake wave field maximum. The laser-driven electron stripping occurs around $t-t_{\mathrm{ref}}\simeq-6\,\mathrm{fs}$, followed by the development of transverse and longitudinal wakefields. The dotted lines indicate the on-axis $|E_y|$  and $|E_x|$ maximas, occurring at $t-t_{\mathrm{ref}}=3.31$, $6.38$ and $9.53\,\mathrm{fs}$ as in Fig.~\ref{fig:fields}(a–f) . These independently observed field maxima in the no-nanoparticle simulations aided in investigating whether the subsequent electron departures at the nanoparticle surface are associated with the plasma wake waves or not. For nanoparticles with radii $R=20.5$ and $51.7\,\mathrm{nm}$, we identify the first observed departure of each stripped electron from the surface of the nanoparticle. The blue and orange symbols in Fig.~\ref{fig:fields}(g–i) distinguish electron departures occurring before and after $t_{\mathrm{ref}}$, respectively. The electron departure, together with the additional stripped charge shown in Fig.~\ref{fig:stripping}(c), shows the second stage of electron stripping associated with the plasma wake wave.

\begin{figure*}[!t]
    \centering
    \includegraphics[width=\textwidth]{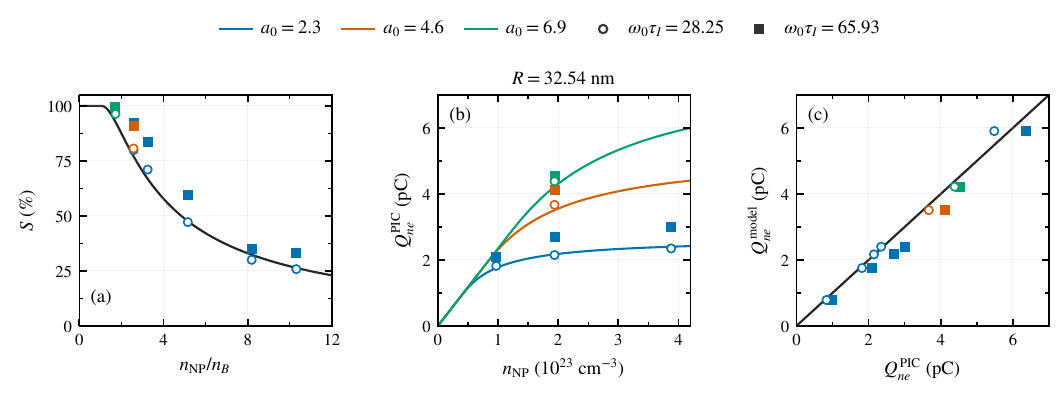}
    \caption{Pulse duration dependence of laser-driven electron
    stripping in vacuum. (a) Stripped electron fraction. (b) Stripped charge for different nanoparticle electron densities at $R=32.54\,\mathrm{nm}$.
    (c) Comparison of model and PIC stripped charge.
    Panels (a) and (c) contain all 14 vacuum simulations, and panel (b) contains the 10 cases at
    $R=32.54\,\mathrm{nm}$.}
    \label{fig:longpulse_appendix}
\end{figure*}

\prlsection{Limitations}
The model treats the ions as a uniform, fixed sphere and the unstripped electrons as a cold, spherical volume in quasistatic force balance. It neglects electron inertia, heating, return, and the evolving field of stripped electrons~\cite{BreizmanArefiev2003,Breizman2005,Arndt2006}, and is not time-dependent. To investigate the effects of these approximations, we studied the second set of vacuum PIC simulations performed with the grid parameters summarized in Table~\ref{tab:grid_parameters} and laser parameters of the \textit{Simulations and methods} section. The seven simulation scans were selected to understand the effects of nanoparticle electron density, radius, and laser amplitude on the stripping produced by multicycle pulses. Using laser amplitudes of $a_0=2.3$, $4.6$, and $6.9$, the multicycle pulses produce more stripped charge across all seven studied cases. The model predicts less for all results with $\omega_0\tau_I=65.93$, as shown in Fig.~\ref{fig:longpulse_appendix}. At fixed $R=32.54\,\mathrm{nm}$ and $a_0=2.3$, the model predicts a stripped charge of $1.75\,\mathrm{pC}$ (77.07\%) at nanoparticle electron density, $n_{\mathrm{NP}}=0.97\times10^{23}\,\mathrm{cm}^{-3}$, while the PIC simulations with $\omega_0\tau_I=28.25$ and $65.93$ produce $1.81\,\mathrm{pC}$ (79.90\%) and $2.1\,\mathrm{pC}$ (92.31\%), respectively. At the higher nanoparticle electron density, $n_{\mathrm{NP}}=3.88\times10^{23}\,\mathrm{cm}^{-3}$, the model predicts $2.39\,\mathrm{pC}$ (26.36\%), while the PIC simulations with $\omega_0\tau_I=28.25$ and $65.93$ produce $2.35\,\mathrm{pC}$ (25.79\%) and $3.01\,\mathrm{pC}$ (33.07\%), respectively. Comparing the PIC results for $\omega_0\tau_I=65.93$ with the model predictions, we see that the increase in stripped charge is greater at higher nanoparticle electron density, even though the corresponding increase in stripped electron fraction is smaller. Electron trajectories in PIC show that multicycle pulses increase the contribution to stripping from electrons in the nanoparticle core; electron return and broadening of the electron momentum distribution are also observed. These benchmarks with multicycle pulses show that the electron stripping process depends on the stripped electron dynamics and nanoparticle core interaction with the different optical cycles of the laser pulse. The model also does not describe the phenomenon of plasma wake waves or its stripped charge, $Q_{pe}$.

\prlsection{Conclusion}
We have shown that an electrostatic model can predict the charge stripped by the laser from the nanoparticle radius, electron density, and laser amplitude. We found that the saturation density $n_B$ given in Eq.~\eqref{eq:model} separates complete from partial stripping, while $Q_\infty$ defines the high-density charge limit. Our PIC simulations showed good agreement with the predicted density, radius, and field scaling of laser-driven stripping in vacuum and helium plasma. We demonstrated why increasing nanoparticle electron density alone eventually produces only small increases in stripped nanoparticle charge. We also showed that laser pulse duration influences electron stripping beyond the predictions of the electrostatic model. Pulses with $\omega_0\tau_I=65.93$ produced more stripped charge than those with $\omega_0\tau_I=28.25$, with the model consistently underpredicting the results for the multicycle pulses. We found that multicycle pulses increased the contribution to stripping from electrons initially located in the nanoparticle core. We also observed electron return and broadening of the electron momentum distribution in the PIC simulations. Together, these observations showed that nanoparticle electron stripping depends on evolving electron dynamics and interactions over successive optical cycles of the laser pulse, which are not captured by the electrostatic model. For experimental applications, our results provide insight into the scaling of nanoparticle charge stripped by the laser and identify the nanoparticle density range beyond which changes in nanoparticle radius or laser amplitude become more or less effective. In helium plasma, we also observed additional electron stripping induced by the plasma wake after the laser-driven electron stripping. Since laser-driven electron stripping produces nanoparticle charge-separation potential that may contribute to electron injection, these results could help inform the selection of nanoparticle parameters for future nanoparticle-assisted laser wakefield acceleration experiments.

\prlsection{Acknowledgments}
This work was supported by the Ministry of Education, Youth and Sports of the Czech Republic through e-INFRA CZ (ID:90254). Portions of this research were carried out at the ELI Beamlines Facility, a European user facility operated by the Extreme Light Infrastructure ERIC.

\prlsection{Data availability}
 Data supporting the findings of this study are available from the corresponding author on a reasonable request. 

\bibliographystyle{apsrev4-2}
\bibliography{references}

\end{document}